\documentclass[10pt]{llncs}
\usepackage{latexsym,amsfonts}
\usepackage{amsmath}
\usepackage{epsfig}
\usepackage{amssymb}
\usepackage{algorithmic}
\usepackage{algorithm}
\usepackage{array}

\newcommand{\A}{\mathcal{A}}
\newcommand{\B}{\mathcal{B}}
\newcommand{\E}{\mathcal{E}}

\newcommand{\I}{\mathcal{I}}
\newcommand{\odd}{\mathcal{O}}
\newcommand{\cost}{\mathsf{cost}}
\newcommand{\OPT}{\mathsf{OPT}}
\newcommand{\Algo}{\mathsf{Algo1}}
\newcommand{\Capacity}{\mathsf{copies}}
\newcommand{\U}{\mathcal{U}}

\title{Popularity at Minimum Cost\thanks{This work was done as part of the DST-MPG partner group ``Efficient Graph Algorithms''.}}
\author{Telikepalli Kavitha\inst{1}, Meghana Nasre\inst{2}, Prajakta Nimbhorkar\inst{3}}
\institute{Tata Institute of Fundamental Research, India. \ \ \ {\sf kavitha@tcs.tifr.res.in} \and Indian Institute of Science, India.\ \ \ {\sf meghana@csa.iisc.ernet.in} \and The Institute of Mathematical Sciences, India. \ \ \ {\sf prajakta@imsc.res.in}}

\begin{document}

\maketitle

\begin{abstract}
We consider an extension of the {\em popular matching} problem in this paper. 
The input to the popular matching problem 
is a bipartite graph $G = (\A \cup \B,E)$, where $\A$ is a set of people, $\B$ is a set of items, and
each person $a \in \A$ ranks a subset of items in an order of preference, with ties allowed.
The popular matching problem seeks to compute a matching $M^*$ between people and items such
that there is no matching $M$ where more people are happier with $M$
than with $M^*$. Such a matching $M^*$ is called a popular matching. 
However, there are simple instances where no popular matching exists. 

\smallskip

Here we consider the following natural extension to the above problem: 
associated with each item $b \in \B$ is a non-negative price
$\cost(b)$, that is,
for any item $b$, new copies of $b$ can be added to the input graph by paying an amount of $\cost(b)$
per copy. When $G$ does not admit a popular matching,
the problem is to ``augment'' $G$ at minimum cost such that the new graph admits a 
popular matching. 
We show that this problem is NP-hard; in fact, it is NP-hard to approximate it within a factor of
$\sqrt{n_1}/2$,
where $n_1$ is the number of people. This problem has a simple polynomial time algorithm when each 
person has a preference list of length at most 2. However, if we consider the problem of
{\em constructing} a graph at minimum cost that admits a popular matching that matches all people, 
then even with preference lists of length 2, the problem becomes NP-hard. 
%
%
On the other hand, when the number of copies of each item is {\em fixed}, we show that
the problem of computing a minimum cost popular matching or deciding that no popular matching exists 
can be solved in $O(mn_1)$ time, where $m$ is the number of edges.
\end{abstract}

\section{Introduction}
\label{intro}

The {\em popular matching} problem deals with matching people to items, where each person ranks a 
subset of items in an order of preference, with ties allowed. The input is a bipartite graph
$G = (\A\cup\B,E)$ where $\A$ is the set of people, $\B$ is the set of items and the edge set 
$E = E_1 \cup \cdots \cup E_r$ ($E_i$ is the set of edges of rank $i$).
For any $a \in \A$, we say $a$ prefers item $b$ to item $b'$ if
the rank of edge $(a,b)$ is smaller than the rank of edge $(a,b')$. If the ranks of
$(a,b)$ and $(a,b')$ are the same, then $a$ is indifferent between $b$ and $b'$.
The goal is to match people with items in an {\em optimal} manner, where the definition of optimality will be 
a function 
of the preferences expressed by the elements of $\A$. 
The problem of computing such an optimal matching is a well studied problem and
several notions of optimality have been
considered so far; for instance,
pareto-optimality \cite{ACMM04}, rank-maximality \cite{IKMMP04}, and fairness. 

One criterion that does not use the absolute values of the ranks  is the notion of
{\em popularity}. Let $M(a)$ denote the item to which a person  $a$ is matched in 
a matching $M$.
We say that a person $a$ \emph {prefers} matching $M$ to $M'$ if
(i)~$a$ is matched in $M$ and unmatched in
$M'$, or (ii)~$a$ is matched in both $M$ and $M'$, and $a$ prefers $M(a)$ to $M'(a)$.

\begin{definition}
$M$ is \emph {more popular than} $M'$, denoted by $M \succ M'$, if the number of people
who prefer $M$ to $M'$ is higher than those that prefer $M'$ to $M$.
A matching $M^*$ is \emph {popular} if there is no matching that is more popular than $M^*$.
\end {definition}

Popular matchings were first introduced by G\"{a}rdenfors~\cite{Gar75} in the context of stable matchings for two-sided preference lists (here both sides of the 
graph $G$ express preferences). In the domain of one-sided preference lists,
popular matchings can be considered to be stable as no majority vote of people
can force a migration to another matching; also, popularity is based on {\em relative} 
ranking rather than the actual ranks used by people. These properties make
 popularity a desirable notion of
optimality in the domain of one-sided preferences.

On the flip side, popularity does not provide a complete answer since 
there exist simple instances that do not admit 
any popular matching. An example is the following: 
let $\A = \{a_1, a_2, a_3\}$,
$\B = \{b_1, b_2,b_3\}$, and each person $a_i$, for $i = 1,2,3$, prefers $b_1$ to $b_2$, and $b_2$ to $b_3$.
Consider the three symmetrical matchings $M_1 = \{(a_1, b_1)$, $(a_2, b_2)$, $(a_3, b_3)\}$,
$M_2 = \{(a_1, b_3)$, $(a_2, b_1)$, $(a_3, b_2)\}$ and
$M_3 = \{(a_1, b_2)$, $(a_2, b_3)$, $(a_3, b_1)\}$.
None of these matchings is popular, since $M_1 \prec M_2$, $M_2 \prec M_3$, and $M_3 \prec M_1$.
Abraham et al. \cite{AIKM05} designed efficient algorithms for determining if a given instance
admits a popular matching and computing one, if it exists.

The fact that popular matchings do not always exist has motivated several extensions to the popular matching problem. 
McCutchen \cite{McC08} considered the
problem of computing a {\em least unpopular} matching; he considered two measures of unpopularity and showed 
that computing a matching that minimized either of these measures is NP-hard.
Kavitha et al. \cite{KMN09} 
generalized the notion of popularity to {\em mixed matchings} or probability
distributions over matchings and showed that a popular mixed matching always exists.

\subsubsection*{Our problem.}
Here we consider another natural generalization to the popular matching problem: {\em augment} the input graph $G$ 
such that the new graph admits a popular matching. Our input consists of $G = (\A\cup\B, E)$ and a function 
$\cost: \B \rightarrow \mathbb{R}^+$,
where $\cost(b)$ for any $b \in \B$ is the cost of making a new copy of item $b$. The set $\B$ is a set of
items, say books or DVDs, and new copies of any $b \in \B$ can be obtained by paying 
$\cost(b)$ for each new copy of $b$. There is no restriction on the number of copies of any item that can be
made. The only criterion that we seek to optimize is the total cost of augmenting $G$.

Going to back to the earlier example on 3 people and 3 items that did not admit a popular matching, 
it is easy to show that by making a new copy of either
$b_1$ or $b_2$, the resulting graph admits a popular matching. In order to minimize the cost, we will make a new copy of
that item in $\{b_1,b_2\}$ which has lower cost.
Our starting graph
$G = (\A\cup\B,E)$ comes for free, every {\em addition} that we make to $G$ comes at a price and our goal is
to make these additions such that  the new graph admits a 
popular matching and the total cost of additions is minimized. We call this the {\em min-cost augmentation} problem.

\subsubsection*{A related problem.}
A related problem is the following: we do not have a starting graph $G$. We are
given a set $\A$ of people and their preference lists over a universe $U$ 
of items where each item $b \in U$ has a price $\cost(b) \ge 0$ associated with it. 
The problem is to ``construct'' 
an input graph $G = (\A\cup\B, E)$ where $\B$ is a multiset of some elements in $U$ 
such that $G$ admits a popular matching
and the cost of constructing $G$, that is, $\sum_{b\in\B} \cost(b)$, is as small as possible.
Here we also have an extra condition
that the popular matching should leave no person unmatched, otherwise we have a trivial solution of $\B = \emptyset$.
We call this problem the {\em min-cost popular instance} problem.

The above problem can also be regarded as a ``gift buying'' problem. Each person in $\A$ has a preference list over
gifts that she would like to receive. The problem is to buy a gift for each person in $\A$
with the total cost as small as possible and 
assign each person a gift such that this assignment is popular. 
That is, there is no reassignment of gifts that makes more people happier. This is the min-cost popular instance problem.

\subsection{Our Results}
We show the following results in this paper:
\begin{enumerate}
\item The min-cost popular instance problem is NP-hard, even when each preference list has length
at most 2 (i.e., every person has a top choice item and possibly, a second choice item).
\item The min-cost augmentation problem has a polynomial time algorithm when each preference list 
has length at most 2 .
\item The min-cost augmentation problem is NP-hard for general lists. In fact, it is NP-hard to approximate
to within a factor of $\sqrt{n_1}/2$, where $n_1$ is the number of people.
\end{enumerate}

All our NP-hardness results hold even when preference lists are derived from a 
{\em master list}. A master
list is a total ordering of the items according to some global objective criterion. Thus if $b_1$ precedes $b_2$
in the master list and if a person $a$ has both $b_1$ and $b_2$ in her list, then it has to be the case that
$b_1$ precedes $b_2$ in $a$'s list.

We would like to contrast the NP-hardness of the min-cost augmentation problem with the problem of
determining a popular matching with variable job capacities \cite{KN09}. In this problem
 the input is a graph 
$G = (\A\cup\B,E)$ where $\A$ is a set of people and $\B$ is a set of jobs, we are also given
a list $\langle c_1,\ldots,c_{|\B|}\rangle$ denoting upper bounds on the capacities of each job. The problem is
to determine if there exists $(x_1,\ldots,x_{|\B|})$ such that for each $i$,
setting the capacity of the $i$-th job to $x_i$, where $1 \le x_i \le c_i$,
enables the resulting graph to admit a popular matching. This problem was shown to be NP-hard in \cite{KN09}; 
however note that 
here the capacity of {\em each} job has an upper bound. Instead, if we only had to maintain an overall upper bound 
on the total increase of capacities  rather than individual upper bounds, a simple polynomial time
algorithm solves this problem \cite{KN09}. 

\smallskip

In the min-cost augmentation problem  recall that there is no upper bound on the amount that we can spend on a
particular item. What we seek to optimize is the overall cost and this problem is NP-hard.
Note that when each item has the same
cost, then this problem can be solved in polynomial  time (using the above algorithm from \cite{KN09}).
However when the costs come from $\{1,2\}$ the problem becomes NP-hard.

\smallskip

The NP-hardness results for the min-cost augmentation/min-cost popular instance problems are 
because the 
number of copies of each of the items need
to be determined
so as to ensure the existence of a popular matching at minimum cost.  Let
$\Capacity(b)$ for any item $b \in \B$ denote the number of copies of item $b$ in our graph $G$.
We now consider
the following problem: each $b \in \B$ has a {\em fixed} number of copies denoted by $\Capacity(b)$ and let the cost of a matching $M$ be 
the sum of costs of items that are matched in $M$ (we have to pay a cost of $k.\cost(b)$ if 
$k$ copies of item $b$ are used in $M$, where $k \le \Capacity(b)$).
Our final result is a polynomial time algorithm
for the {\em min-cost popular matching} problem which we define below.

\begin{itemize}
\item[$\ast$] The {\em min-cost popular matching} problem is to determine if $G$ admits a 
popular matching or not and if so, to compute the one with minimum cost. We show that this problem can be 
solved in $O(mn_1)$ time, where $m$ is the number of edges and $n_1$ is the number of people. Manlove and Sng
considered this problem without costs in the context of House Allocation. There items were called houses
and copies of items as in our case were represented using capacities for houses. They called
it Capacitated House Allocation with Ties (CHAT) and the problem was to determine if
$G$ admits a popular matching or not, and if so, to compute one. Manlove and Sng \cite{MS06}
showed an $O((\sqrt{C}+n_1)m)$ algorithm for the CHAT problem, where $C$ is the sum of 
capacities of
all items. 
Thus, our algorithm improves upon the
algorithm in \cite{MS06}.
\end{itemize}

\subsection{Background}

Abraham et al. \cite{AIKM05} considered the problem of determining if a given graph $G = (\A\cup\B,E)$
admits a popular matching or not, and if so, computing one. They also gave a structural characterization
of graphs that admit popular matchings. Section~\ref{prelims} outlines this characterization and the algorithm
that follows from it.

Subsequent to this work,
several variants of the popular matchings problem have been considered.
One line of research has been on generalizations of the popular matchings problem
while the other direction has been to deal with instances that do not admit any popular
matchings. The generalizations include the capacitated version studied by Manlove
and Sng \cite{MS06}, the weighted version studied by Mestre \cite{Mes06} and random popular matchings studied
by Mahdian \cite{Mah06}. Kavitha and Nasre \cite{KN08} as well as McDermind and Irving \cite{MI09} independently studied 
the problem of computing an {\em optimal} popular matching for strict instances where the notion of optimality
is specified as a part of the input. Note that they also considered the min-cost popular matchings
but in this version the costs are associated with edges whereas in our problem costs are associated with
items.
As described earlier, the line of research pursued for instances
that do not admit popular matchings includes the NP-hardness of least unpopular matchings
\cite{McC08}, the existence and algorithms for popular mixed matchings \cite{KMN09} and NP-hardness
of the popular matchings problem with variable job capacities \cite{KN09}.

\paragraph{Organization of the paper.} Section~\ref{prelims} discusses preliminaries. 
Section~\ref{min-cost-pop-inst}
shows that the min-cost popular instance problem is NP-hard. 
Section~\ref{augment} has our results
for the min-cost augmentation problem and Section~\ref{min-cost-pop-mat} has our algorithm for the 
min-cost popular matching problem.

\section{Preliminaries}
\label{prelims}
We review the characterization of popular matchings given in
\cite{AIKM05}. 
Let $G_1 = (\A \cup \B, E_1)$ be the graph containing only rank-1 edges. Then
\cite[Lemma~3.1]{AIKM05} shows that a matching $M$ is popular in $G$ only if
$M \cap E_1$ is a maximum matching of $G_1$.
Maximum matchings have the following important properties, which we use throughout the rest of the paper.

$M \cap E_1$ defines a partition of $\A \cup \B$ into three disjoint sets:
a vertex $u \in \A \cup \B$ is \emph {even} (resp. \emph {odd}) if there is an even (resp. odd) length alternating path in
$G_1$ (w.r.t. $M \cap E_1$) from an unmatched vertex to $u$.
Similarly, a vertex $u$ is \emph {unreachable} if there is no alternating path from an unmatched vertex to $u$.
Denote by $\mathcal E$, $\mathcal O$ and $\mathcal U$ the sets of even, odd, and unreachable vertices, respectively.

\begin{lemma}[Gallai-Edmonds Decomposition]
\label{lemma:node-classification}
Let $\mathcal E$, $\mathcal O$ and $\mathcal U$ be the sets of vertices defined by $G_1$ and
$M \cap E_1$ above. Then
\begin {itemize}
\item [(a)] $\mathcal E$, $\mathcal O$ and $\mathcal U$ are pairwise disjoint, and independent of the maximum matching $M \cap E_1$.
\item [(b)] In any maximum matching of $G_1$, every vertex in $\mathcal O$ is matched with a vertex in
$\mathcal E$, and every vertex in $\mathcal U$ is matched with another vertex in $\mathcal U$.
The size of a maximum matching is $|\mathcal O| + |\mathcal U|/2$.
\item [(c)] No maximum matching of $G_1$ contains an edge between a vertex in $\mathcal O$ and a vertex 
in $\mathcal O \cup \mathcal U$.
Also, $G_1$ contains no edge between a vertex in $\mathcal E$ and a vertex in $\mathcal E \cup \mathcal U$.
\end {itemize}
\end{lemma}

Since every maximum cardinality matching in $G_1$ matches all vertices $u \in \mathcal O \cup \mathcal U$,
these vertices are called {\em critical} as opposed to vertices $u \in \mathcal E$ which are called 
{\em non-critical}. Using this partition of vertices, the following definitions can be made.
\begin{definition}
For each $a \in \A$, define $f(a)$ to be the set of  top choice items for $a$.
Define $s(a)$ to be the set of $a$'s most-preferred {\em non-critical} items in $G_1$.
\end{definition}
\begin{theorem}[from \cite{AIKM05}]
\label{thm:pop-mat}
A matching $M$ is popular in $G$ iff 
(i) $M \cap E_1$ is a maximum matching of $G_1 = (\A\cup\B, E_1)$, and 
(ii) for each person $a$, $M(a) \in f(a) \cup s(a)$.
\end{theorem}
The algorithm for solving the popular matching problem is now straightforward:
each $a \in \A$ determines the sets $f(a)$ and $s(a)$. A
matching that is maximum in $G_1$ and that matches each $a$ to an item  in
$f(a) \cup s(a)$ needs to be determined. If no such matching exists, then $G$ does not admit a popular
matching.

\section{Min-cost Popular Instance}
\label{min-cost-pop-inst}
In this section we consider the min-cost popular instance problem.
Our input is a set $\A$ of people where each $a \in \A$ has a preference list over items in
a universe $U$, where each item $b \in U$ has a price $\cost(b) \ge 0$.
The problem is to ``construct'' a graph $G$ or equivalently, set suitable values for $\Capacity(b)$
where $b \in U$, in order to ensure that the resulting graph $G$
admits a popular matching that matches all $a \in \A$, 
at the least possible cost.

We will show that the above problem is NP-hard by showing a reduction from the monotone 1-in-3 SAT 
problem to this problem.
The monotone {1-in-3} SAT problem is a variant of the 
3SAT problem where each clause contains exactly 3 literals and no literal 
appears in negated form. The monotone 1-in-3 SAT problem asks
if there exists a satisfying assignment to the variables such that each clause has exactly 1 literal
set to true.  This problem is NP-hard \cite{Sch78}.

Let $\I$ be an instance of the monotone 1-in-3 SAT problem. 
Let $C_1,\ldots,C_m$ be the clauses in $\I$ and let $X_1,\ldots, X_n$
be the variables in $\I$. We construct from $\I$ an instance of the 
min-cost popular instance problem as follows:

Corresponding to each clause $C_i = (X_{j_1} \vee X_{j_2} \vee X_{j_3})$, we have 9 people
$A_i = \{a^i_1\ldots,a^i_9\}$. Their preference lists are shown in Fig.~\ref{fig:clausegadget-mc-pop-inst}.
In this case every person has a preference list of length 2, that is a top item followed by
a second choice item. For instance, $a_1^i$ treats item $u_{j_1}$ as its rank-1 item and
item $u_{j_2}$ as its rank-2 item.

\begin{figure}[ht]
\begin{minipage}[b]{0.3\linewidth}
\centering
\begin{equation*}
  \setlength{\arraycolsep}{0.5ex}\setlength{\extrarowheight}{0.25ex}
 \begin{array}{@{\hspace{1ex}}c@{\hspace{1ex}}||@{\hspace{1ex}}c@{\hspace{1ex}}|@{\hspace{1ex}}c@{\hspace{1ex}}|@{\hspace{1ex}}c@{\hspace{1ex}}} 
    a_{1}^i & u_{j_1} \ & u_{j_2} & \\[.5ex] \hline
    a_{2}^i & u_{j_2} \ & u_{j_3} & \\[.5ex] \hline
    a_{3}^i & u_{j_1} \ & u_{j_3} & \\[.5ex] 
  \end{array}
\end{equation*}
\end{minipage}
\begin{minipage}[b]{0.3\linewidth}
\centering
\begin{equation*}
  \setlength{\arraycolsep}{0.5ex}\setlength{\extrarowheight}{0.25ex}
  \begin{array}{@{\hspace{1ex}}c@{\hspace{1ex}}||@{\hspace{1ex}}c@{\hspace{1ex}}|@{\hspace{1ex}}c@{\hspace{1ex}}|@{\hspace{1ex}}c@{\hspace{1ex}}}
    a_{4}^i & u_{j_1} \ & p_1^i& \\[.5ex] \hline
    a_{5}^i & u_{j_2} \ & p_2^i& \\[.5ex] \hline
    a_{6}^i & u_{j_3} \ & p_3^i& \\[.5ex] 
\end{array}
\end{equation*}
\end{minipage}
\begin{minipage}[b]{0.3\linewidth}
\centering
\begin{equation*}
  \setlength{\arraycolsep}{0.5ex}\setlength{\extrarowheight}{0.25ex}
  \begin{array}{@{\hspace{1ex}}c@{\hspace{1ex}}||@{\hspace{1ex}}c@{\hspace{1ex}}|@{\hspace{1ex}}c@{\hspace{1ex}}|@{\hspace{1ex}}c@{\hspace{1ex}}}
    a_{7}^i & p_1^i \ & q^i& \\[.5ex] \hline
    a_{8}^i & p_2^i \ & q^i& \\[.5ex] \hline
    a_{9}^i & p_3^i \ & q^i& \\[.5ex] 
\end{array}
\end{equation*}
\end{minipage}
\caption{The preference lists of people corresponding to the $i$-th clause in $\cal I$.}

\label{fig:clausegadget-mc-pop-inst}
\end{figure}

The items $u_{j_1},u_{j_2},u_{j_3}$ are called {\em public} items and the items $p_1^i,p_2^i,p_3^i$, and $q^i$
are called {\em internal} items. The internal items appear only on the preference lists of the people of $A_i$
while the public items also appear on the preference lists of people in $A_i$ as well as outside $A_i$.
The public item $u_j$ corresponds to the variable $X_j$. In every clause $C_i$ that $X_j$ 
belongs to, the item $u_j$ appears in the preference lists of some of the people in the set $A_i$  
as shown in Fig.~\ref{fig:clausegadget-mc-pop-inst}.

\medskip

The set $\A$ of people in our instance is $\cup_i A_i$.
The universe $U$ of all items is the union of $\{u_1,\ldots,u_n\}$ (the $n$ public items) and 
the set $\cup_i \{p^i_1,p^i_2,p^i_3,q^i\}$ of all the internal items. 
It remains to describe the costs of the items.
For each $i$, the cost of each $p_t^i$ for $t = 1,2,3$, is 1 unit, while the cost of $q^i$ is zero units.
The cost of each $u_j$, for $j = 1,\ldots,n$, is 3 units. 

Recall that our problem is to determine a set $\B$ of items with suitable copies so that the graph 
$(\A\cup\B,E)$
admits a popular matching that matches all $a \in \A$ 
and we want to do this at the least possible cost. 
We first show the following lemma.

\begin{lemma}
\label{lemma:new}
Any instance $(\A \cup \B, E)$ that admits a popular matching that matches all $a \in \A$ has cost at least $14m$.
\end{lemma}
\begin{proof}
Let us focus on the set $A_i$ of people corresponding to clause $C_i$. The preference lists of people in $A_i$
are shown in Fig.~\ref{fig:clausegadget-mc-pop-inst}.
Since the cost of each item on the lists of $a^i_1,a^i_2,a^i_3$ is 3, 
we have to spend 9 units to buy an item each 
for these 3 people (since we seek an instance where all the persons get matched). 
People $a_4^i,a_5^i,a_6^i$ have a unit cost item in their preference lists 
(items $p_1^i,p_2^i,p_3^i$, respectively). 
Thus, we have to spend 3 units to buy an item each for these 3 people.
Finally, $a_7^i,a_8^i,a_9^i$ have a cost 0 item, i.e.
$q^i$, in their preference lists. Hence, we can get $q^i$ with $\Capacity(q^i) = 3$ for a cost of 0. 
Summarizing, we
need to spend at least $9 + 3 + 0 = 12$ units for the people in $A_i$.

However, it is not possible to spend just 12 units for the people in $A_i$. 
This is because, in the first place,
we are forced to have non-zero copies for at least 2 of the 3 items in $\{u_{j_1},u_{j_2},u_{j_3}\}$ and
if we seek to match $a_4^i$ to $p_1^i$ while $u_{j_1}$ is around, then $p_1^i$ is $a_4^i$'s second choice item. 
Since $p_1^i$ is $a_7^i$'s top choice item, we also have
to match $a_7^i$ to $p_1^i$ since a popular matching has to be a maximum cardinality matching on rank-1 edges
(see Theorem~\ref{thm:pop-mat}). Thus, it is not possible to match $a_7^i$ to $q^i$ in a popular matching 
while $p_1^i$ 
gets matched  to $a_4^i$ who regards this item as a second choice item 
because $u_{j_1}$ is around.\footnote{A matching that 
contains the 3 edges $(a_1^i,u_{j_1}^i), (a_4^i,p_1^i),(a_7^i,q^i)$ cannot be popular since by promoting $a_7^i$ from 
$q^i$ to $p_1^i$, 
and $a_4^i$ from $p_1^i$ to $u_{j_1}$, and leaving $a_1^i$ unmatched, we get a more popular matching.} 
Thus we have the following options:
\begin{itemize}
\item[(i)] set $\Capacity(u_{j_1}) = 0$ and then match $a_4^i$ to $p_1^i$ and $a_7^i$ to $q^i$
\item[(ii)] match both $a_4^i$ and $a_7^i$ to $p_1^i$ by setting $\Capacity(p_1^i) = 2$
\item[(iii)] add one more copy of $u_{j_1}$ and set $\Capacity(p_1^i) = 0$ and thus match $a_4^i$ to $u_{j_1}$ and $a_7^i$ to $q^i$  
\end{itemize}

It is not possible to have option (i) for all the $u_j$'s 
since we are forced to have non-zero copies for at least 2 of the 3 items in $\{u_{j_1},u_{j_2},u_{j_3}\}$.
Note that (ii) is always better than (iii) (cost of 2 units vs cost of 3 units). Hence it is always cheaper 
to match $a_4^i,a_5^i,a_6^i$ to $p_1^i,p_2^i,p_3^i$ respectively than to any of $u_{j_1},u_{j_2},u_{j_3}$.

Thus, when we match $a_4^i,a_5^i,a_6^i$ to $p_1^i,p_2^i,p_3^i$ respectively,
at least 2 of these 3 people are getting matched to their second choice items. 
Hence, at least
2 out of the 3 people among  $a_7^i,a_8^i,a_9^i$
will also have to be matched to their top choice items
in order to ensure that the resulting matching is popular. This implies
a cost of at least $9 + 3 + 2 = 14$ for $A_i$. 

This holds for each $A_i$, where $1 \le i \le m$. Since the cost is at least 14 per clause, it amounts to 
at least $14m$ in total for all the clauses. \qed 
\end{proof}

The following lemma establishes the correspondence between the instance $\I$ of monotone 1-in-3-SAT and the min-cost popular
instance that we defined.

\begin{lemma}
There exists an instance $(\A \cup \B, E)$   with cost $14m$  that admits a popular matching that matches all $a \in \A$
iff there exists a 1-in-3 satisfying assignment for $\I$.
\end{lemma}
\begin{proof}
We know from Lemma~\ref{lemma:new} that any instance $(\A \cup \B, E)$
that admits a popular matching that matches all $a \in \A$ has a cost of at least
$14m$. What we need to show here is that $(\A \cup \B, E)$ has cost $14m$ if and only if the
1-in-3-SAT instance $\I$ is a ``yes'' instance, that is,  there is a true/false assignment to the variables
$X_1,\ldots,X_n$ such that each clause has exactly 1 literal set to true (and thus 2 literals set to false).

Suppose $\I$ admits such an assignment. We now show how to construct a set $\B$ of cost $14m$  such that the
instance $(\A \cup \B, E)$ admits a popular matching that matches all $a \in \A$.
If $X_i = true$ then set $\Capacity(u_i) = 0$, else $\Capacity(u_i)$ will be set to a suitable 
strictly positive value.

Since the setting of true/false values to $X_i$'s is a satisfying assignment, 
every clause has two literals set to false and 1 set to true. 
Let clause $C_i$ be $(X_{j_1}\vee X_{j_2} \vee X_{j_3})$.
Thus there is 1 variable
$X_{j_k}$ in $\{X_{j_1},X_{j_2},X_{j_3}\}$ that has been set to true. By our definition of copies of every item,
the corresponding $u_{j_k}$ has 0 copies. Hence the people in the set $A_i$ can be matched as follows:
\begin{itemize}
\item $a^i_1,a^i_2,a^i_3$ get matched to the 2 items in $\{u_{j_1},u_{j_2},u_{j_3}\} \setminus \{u_{j_k}\}$
by having 2 copies of one of the lower indexed item and 1 copy of the higher indexed item for these 3 people.
\item $p^i_{k}$ becomes $a^i_{k+3}$'s top choice item (since $u_{j_k}$ does not exist in the graph now) and
hence we can now match $a_{k+3}^i$ to $p^i_k$ and $a_{k+6}^i$ to $q^i$.
\end{itemize}
This way we spend only $9 + 3 + 2 = 14$ units
for the people in $A_i$ and each person $a$ has an item in $f(a)\cup s(a)$ to be matched to. 
Since every clause in $\I$ has exactly 1 variable set to true and 2 set to false,
we achieve a cost of 14 for each set $A_i$. 
This shows that we can construct a set $\B$ of cost $14m$
such that $(\A \cup \B, E)$ admits a popular matching that matches all $a \in \A$.

\medskip

To show the other direction, let us set the true/false values of variables in $\I$ as follows:
for each $j = 1,\ldots,n$ set $X_{j}=true$ if and only if $\Capacity(u_{j}) = 0$.
We need to show that such an assignment
sets exactly 1 variable in each clause to be true.

Let us consider any clause $C_i = (X_{j_1} \vee X_{j_2} \vee X_{j_3})$.
Among the 3 items $u_{j_1},u_{j_2},u_{j_3}$ that correspond to these 3 variables
we need at least 2 items to have non-zero copies so as to match all the 3 people $a^i_1,a^i_2,
a^i_3$. Thus, our true/false assignment does not set more than 1 variable per clause to true.

We now need to show  that there is at least 1 item in $\{u_{j_1},u_{j_2},u_{j_3}\}$ with zero copies.
This is where we will use the hypothesis that we can construct
$(\A \cup \B, E)$ of cost $14m$ that admits a popular matching that matches all $a \in \A$.
It follows from the proof of Lemma~\ref{lemma:new}
that each set $A_i$ of people corresponding to a clause needs a cost of at least 14. Since the total cost is only $14m$ and there
are $m$ clauses, this
implies that we have to spend exactly 14 per clause. In other words, the items for the 9 people of each $A_i$
have to be bought using only 14 units.

If all the 3 items in $\{u_{j_1},u_{j_2},u_{j_3}\}$
have non-zero copies, then this implies the
cost of items for all the 9 people in $A_i$ will be $9 + 3 + 3 = 15$ since when each $u_{j_k}$ has at least one copy,
then the $u_{j_k}$'s become top choice items for $a^i_4,a^i_5,a^i_6$, respectively and thus
$p_1^i,p_2^i,p_3^i$ become their second choice items.
This forces us to match each of $a_7^i,a_8^i,a_9^i$ to their top choice items (that is, $p_1^i,p_2^i,p_3^i$,
respectively) since a popular matching has to be a maximum cardinality matching on rank-1 edges.
However, we are given that we can spend only 14 units per $A_i$; 
thus it has to be the case that there exists at least 1 item in $\{u_{j_1},u_{j_2},u_{j_3}\}$ 
which has zero copies. This finishes the proof of this lemma. \qed

\end{proof}

Note that the preference lists of all the people in our instance $G$ are strict and of length
at most 2. Also, the preference lists are drawn from a {\em master list}.
We have thus shown the following theorem.
\begin{theorem}
\label{thm3}
The min-cost popular instance problem is NP-hard, even when each preference list has length at most 2. 
Further, the hardness holds even when the preference lists are derived from a master list.
\end{theorem}
  
\section{Min-cost Augmentation}
\label{augment}
In this section we show various results for the min-cost augmentation problem. Recall that the
input here is a graph $G = (\A\cup\B,E)$ where each item $b \in \B$ has a non-negative $\cost(b)$
associated with it. The problem is to determine how to make extra copies of items in $\B$
so that the resulting graph admits a popular matching and the cost of the extra copies is minimized.

Unlike the min-cost popular instance problem, the above problem admits a simple polynomial time algorithm
when each $a \in \A$ has a preference list that is strict and of length at most 2. 
We describe this algorithm below.
We assume throughout this section that we add at the end of each $a$'s preference list a dummy item called 
the {\em last item} $\ell_a$, where $a$ being matched to $\ell_a$ amounts to $a$ being left unmatched.

\subsection{Preference lists of length 2}
For any $a \in \A$, $a$'s preference list consists of a top choice item (let us use $f_a$ to denote this item),
and possibly a second choice item (let us use $z_a$ to denote this item) and then of course, 
the last item $\ell_a$
that we added for convenience. Let $G_1$ be the graph $G$ restricted to rank-1 edges.
Let the graph $G' = (\A\cup\B,E')$, where $E'$ consists of 
\begin{itemize}
\item all the top ranked edges $(a,f_a)$: one such edge for each $a \in \A$, and 
\item the edges $(a,s_a)$, where $a$ is {\em even} in $G_1$ 
and $s_a$ is $a$'s most preferred item that is {\em even} in $G_1$. Thus $s_a= z_a$ 
when $z_a$ is nobody's top choice item, else
$s_a = \ell_a$. 
\end{itemize}

It follows from Theorem~\ref{thm:pop-mat} that $G$ admits a popular matching if and only if $G'$ admits
an $\A$-perfect matching.
We assume that $G$ does not admit a popular matching and we have to decide now which
items should be duplicated and how many extra copies should be made.
Since
$G'$ does not admit a popular matching, there exists a set $S$ of people
such that the neighborhood $N(S)$ of $S$ in $G'$ satisfies $|N(S)| < |S|$.
Let $S$ denote a minimal such set of people. It is easy to see that every 
$a \in S$ must be even in $G_1$. Thus, for each $a \in S$, 
the edge $(a,s_a)$ belongs to $G'$ and it must be that
$s_a = z_a$. Otherwise $s_a = \ell_a$ and since no vertex in $\A$
other than $a$ has an edge to $\ell_a$, such an $a$ will be always matched
in any maximum cardinality matching in $G'$. Hence, such an $a$ cannot belong to
$S$ due to its minimality. Further note that for any such minimal set $S$,
the set $N(S)$ is a set of items that are all {\em odd} in the graph $G'$ with
respect to a maximum cardinality matching in $G'$.

Since $s_a = z_a$ for every $a \in S$, and the preference lists are of length
at most 2, there are no items sandwiched between $f(a)$ and $s(a)$ in
$a$'s preference list for every $a \in S$. Thus, in order to ensure that these
people get matched in any popular matching, we need to make extra copies
of items in $N(S)$ or equivalently items that are {\em odd} in
the graph $G'$. Our algorithm precisely does this and in order to get
a min-cost augmentation, it chooses the odd item in $G'$ 
which has least cost. The steps
of our algorithm are described in Algorithm~\ref{algo:min-cost-aug-length2}.

\begin{algorithm}
\begin{algorithmic}[1]
\STATE Construct the graph $G'  = (\A\cup\B, E')$ where $E' = \{(a,b): a\in\A, b \in f(a)\cup s(a)\}$.
\STATE $H_0 = G$, $H_0' = G'$, Let $M_0$ denote a maximum cardinality matching in $H_0'$.
\STATE $i=0$.
\WHILE {$M_i$ is not $\A$-complete matching}
\STATE Partition the vertices into {\em odd ($\odd$), even ($\E$), unreachable ($\U$)} w.r.t. $M_i$.
\STATE Let $b$ denote the cheapest item in $\B \cap \odd$.
\STATE Set $\Capacity(b) = \Capacity(b)+1$.
\STATE Construct the graph $H_{i+1}'$ corresponding to $H_{i+1}$ and update $M_{i+1}$ to be a maximum cardinality matching in $H_{i+1}'$.
\STATE $i= i+1$.
\ENDWHILE
\STATE Output the graph $H_i$.
\end{algorithmic}
\caption{Min-cost augmentation for strict lists of length at most 2.}
\label{algo:min-cost-aug-length2}
\end{algorithm}

Our algorithm maintains the invariant that no person $a$ changes her $s$-item
due to the increase in copies. This is because we ensure that
no top choice item $b$ ever becomes {\em even} in $H_i^1$, the
graph $H_i$ restricted to rank-1 edges. Note that the set of {\em odd}
items in $H_i$ is identified by constructing alternating paths from
a person who is unmatched in $H_i$ and every item $b$ that appears 
on such a path is always {\em odd}.
Further, our duplications ensure that the total number of copies
of an item $b$ in any augmented instance $H_i$ is bounded by the
degree of $b$ in $G'$.  In the case of a top choice item $b$, 
the degree of $b$ in $G'$ is equal to
the degree of $b$ in $G_1$, the graph $G$ restricted to rank-1 edges. 
Thus, even with the extra copies, a top choice item remains critical
in the augmented graph restricted to rank-1 edges. This implies
that for every person, the most preferred {\em even} item in the augmented graph
restricted to rank-1 edges (i.e., its $s$-item) remains unchanged.

We note that the above claim also implies that in every iteration
of the while loop in Step~$4$ of our algorithm, the size
of the maximum cardinality matching increases by $1$, that is,
$|M_{i+1}| = |M_{i}|+1$. Therefore, the while loop terminates
in $k = |\A| - |M_0|$ iterations. Since $k$ is bounded
by $n_1$, the number of applicants in $G$, the running
time of our algorithm is $O(n_1^2)$.
It is clear that the
graph $H_i$ returned by the algorithm admits an $\A$-complete
matching the graph $H_i'$ and hence admits a popular matching.
To see that the instance returned is a min-cost instance, observe
that there is no alternating path between an item $b$ which got
duplicated in our algorithm and an item $b'$ whose cost is strictly
smaller than the $\cost(b)$. Otherwise in the iteration when $b$ was {\em odd},
so was $b'$ and it would have been picked up by our algorithm.

We can therefore conclude the following theorem.

\begin{theorem}
The min-cost augmentation problem with strict preference lists of length at most 2
can be solved in $O(n_1^2)$ time.
\end{theorem}

\subsection{Hardness for the general case}
We now show that the min-cost augmentation problem in the general case 
is NP-hard. The reduction is again from the monotone 1-in-3 SAT
problem (refer to Section~\ref{min-cost-pop-inst}). 
Let $\I$ be an instance of the monotone 1-in-3 SAT problem. 
Let $C_1,\ldots,C_m$ be the clauses in $\I$ and let $X_1,\ldots, X_n$
be the variables in $\I$. We construct from $\I$ an instance of the 
min-cost augmentation problem as follows. 

Let $C_i$ be $(X_{j_1} \vee X_{j_2} \vee X_{j_3})$.
Corresponding to this clause we have  6 people
$A_i = \{a_1^{i}, a_2^{i}, a_3^{i}, a_4^{i}, a_5^{i}, a_6^{i}\}$ and 
3 internal items $D_i = \{p_i, q_i, r_i\}$. In addition we have public items 
$u_{j_1}, u_{j_2}, u_{j_3}$ which belong to preference lists of people in $A_i$ and
whenever $X_j$ occurs in a clause $C_i$, the item $u_j$ will belong to the preference lists of 
some people in $A_i$. The public items have unit cost whereas each internal item $b \in D_i$ has cost 2.
The preference lists of the people in $A_i$
are shown in Fig.~\ref{fig:clausegadget-mc-aug}. 

 \begin{figure}[ht]
\begin{minipage}[b]{0.44\linewidth}
 \centering
\begin{equation*}
  \setlength{\arraycolsep}{0.5ex}\setlength{\extrarowheight}{0.25ex}
  \begin{array}{@{\hspace{1ex}}c@{\hspace{1ex}}||@{\hspace{1ex}}c@{\hspace{1ex}}|@{\hspace{1ex}}c@{\hspace{1ex}}|@{\hspace{1ex}}c@{\hspace{1ex}}|@{\hspace{1ex}}c@{\hspace{1ex}}|@{\hspace{1ex}}c@{\hspace{1ex}}}
    a_{1}^i & p_i \ & u_{j_1} \ & q_i \ \\[.5ex] \hline
    a_{2}^i & p_i \ & u_{j_2} \ & q_i \ \\[.5ex] \hline
    a_{3}^i & p_i \ & u_{j_3} \ & q_i \ \\[.5ex] 
\end{array}
\end{equation*}
\end{minipage}
\begin{minipage}[b]{0.44\linewidth}
\centering
\begin{equation*}
  \setlength{\arraycolsep}{0.5ex}\setlength{\extrarowheight}{0.25ex}
  \begin{array}{@{\hspace{1ex}}c@{\hspace{1ex}}||@{\hspace{1ex}}c@{\hspace{1ex}}|@{\hspace{1ex}}c@{\hspace{1ex}}|@{\hspace{1ex}}c@{\hspace{1ex}}}
    a_{4}^i & r_i \ & u_{j_1} & \\[.5ex] \hline
    a_{5}^i & r_i \ & u_{j_2} & \\[.5ex] \hline
    a_{6}^i & r_i \ & u_{j_3} & \\[.5ex] 
\end{array}
\end{equation*}
\end{minipage}
\caption{Preference lists of the 6 people in $A_i$}
\label{fig:clausegadget-mc-aug}
\end{figure}

The set $\B$ of items
is the union of $\cup_{i=1}^m D_i$ (the set of all the internal items) and $\{u_1,\ldots,u_n\}$
(consisting of all the public items, where vertex $u_j$ corresponds to the $j$-th variable $X_j$).
The set $\A$ of people is the union of $\cup_{i=1}^m A_i$ and $\{x_1,\ldots,x_n\}$,
where the vertex $x_j$ corresponds to the variable $X_j$. 
The preference list of each $x_j$ is of length 1, it consists of the item $u_j$.

\paragraph{\bf $G$ has no popular matching.}
It is easy to see that the graph $G$ described above does not admit any popular
matching. To see this, first note that each public item $u_j$ is a unique rank-1 item 
for exactly one applicant $x_j$. 
Hence when every item has a single copy, these public items are unreachable or critical 
in $G_1$ (the subgraph of rank-1 edges in $G$). 
Now let us consider the people in $A_i$: for each $a_t^i\in \{a_1^i, a_2^i, a_3^i\}$,
we have $f(a_t^i) = \{p_i\}$ and $s(a_t^i) = \{q_i\}$.
Since there are only 2 items $p_i, q_i$ for the 3 people $a_1^i, a_2^i, a_3^i$ to be matched to in any
popular matching, $G$ does not admit a popular matching. 

Let $\tilde{G}$ be a min-cost instance  such that
$\tilde{G}$ admits a popular matching. 
We now state the following lemma that establishes the reduction.

\begin{lemma}
\label{lemma5}
$\tilde{G}$ has cost at most $m$ iff there exists a 1-in-3 satisfying 
assignment for the instance $\I$.
\end{lemma}
\begin{proof}
Assume that there exists a 1-in-3 satisfying assignment for $\I$.
For each $j$, let $c_j$ denote the number of clauses in which $X_j$
appears. We will set the copies of the items in the following manner: 
the copies of the internal items remain the same, i.e., $\Capacity(b) = 1$ for each $b \in \cup_i D_i$
and the copies of 
the public items are set as follows.

For each $j$, where $1 \le j \le n$ do:
\begin{itemize}
\item[$\bullet$]  if  $X_j = true$, then set $\Capacity(u_j) = 1 + c_j$
\item[$\bullet$]  else $\Capacity(u_j)$ remains 1.
\end{itemize}

Let us determine the cost of this augmentation. For every $X_j$ that is true,
we pay a cost of $c_j\cdot 1 = c_j$ and for $X_j$ that is false, we pay nothing.
Since each clause has exactly one variable set to true, we have:
$\sum_{j: X_j = true} c_j = m$.
Thus the cost of our augmentation is $m$.

We now show that the graph $\tilde{G}'$ admits an $\A$-complete matching 
(the edges in $\tilde{G}'$ are $(a,b)$ where $b \in f(a)\cup s(a)$). 
\begin{itemize}
\item Consider the people $x_1,\ldots,x_n$.
Each $x_j$ gets matched to her $f$-item $u_j$.
\item Consider the people in $A_i$. We know that
exactly one amongst $u_{j_1}, u_{j_2}, u_{j_3}$ has more than one copies
(since the number of copies was based on a satisfying
assignment for 1-in-3 SAT). If $\Capacity(u_{j_k}) > 1$, then $a_k^i$ gets matched to $u_{j_k}$
and the 2 people in $\{a_1^i, a_2^i, a_3^i\}\setminus\{a_k^i\}$ get matched to $p_i$ and $q_i$.
Finally, $a_{k+3}^i$ gets matched to her top choice item $r_i$ whereas the 2 people in 
$\{a_4^i, a_5^i, a_6^i\}\setminus\{a_{k+3}^i\}$
get matched to their last items (their most preferred even item in $G_1$). 
\end{itemize}

To prove the other direction, assume that the cost of $\tilde{G}$ is $m$.
We now translate this into truth values for variables in $\I$.
If $\Capacity(u_j) > 1$ in $\tilde{G}$, then set variable $X_j = true$, else set $X_j = false$.
We need to show that this is a 1-in-3 satisfying assignment for $\I$.

Since the cost of adding one copy of any item is at least 1,
we need to pay at least 1 unit per clause in order to match the people in $A_i$. Thus, we need to pay at 
least $m$ to get a graph that admits a popular matching. 
However, we are given that with a cost of exactly $m$, the graph $\tilde{G}$ that admits a popular 
matching. 
Hence, the copies of items have been added such that exactly 1 unit has been spent per clause.

Spending 1 unit has allowed all the people in $A_i$, for each $i$, to have
enough items to match themselves to in $\tilde{G}'$. Consider the items that occur in the preference 
lists of
people in $A_i$ (refer to Fig.~\ref{fig:clausegadget-mc-aug}).
Since the cost of each internal item is 2 and we cannot afford a cost of 2 for any clause, 
it has to be the case that $\Capacity(u)> 1$ for some $u \in \{u_{j_1},u_{j_2},u_{j_3}\}$.
Thus, we have at least 1 true variable per clause in $\I$.

We now have  to show that there is exactly 1 true variable per clause in $\I$.
The point to note is that $\Capacity(u) > 1$ for any public item $u$ implies that $u$ is non-critical in 
$\tilde{G}_1$.
This changes the {\em most preferred even} item in $\tilde{G}_1$ for some people. That is,
suppose $k$ items in $\{u_{j_1},u_{j_2},u_{j_3}\}$ have more than 1 copies. 
Then, we have $k$ non-critical items in $\{u_{j_1},u_{j_2},u_{j_3}\}$ and so
we have $k$ people in $\{a_4^i, a_5^i, a_6^i\}$ satisfying the following:
$a$'s {\em most preferred even} item in $\tilde{G}_1$ is no longer 
the last resort item $\ell_a$, it is now the non-critical public item that is second in $a$'s preference list.

Observe that one person in  $\{a_4^i, a_5^i, a_6^i\}$
can be matched to her top choice item $r_i$. However, to match the second person  we need to spend
another unit.
In the first place, we have already spent 1 unit to add an extra copy of some $u_{j_k}$ to match all the people
in $\{a_1^i, a_2^i, a_3^i\}$.
With more than one item in $\{u_{j_1},u_{j_2},u_{j_3}\}$ non-critical in $\tilde{G}_1$, 
we have pay at least 2 units for the people in $A_i$. This contradicts the fact that we spent
exactly 1 unit for the people in $A_i$.
Hence there is exactly 1 true variable per clause in $\I$.
\qed
\end{proof}

We can now conclude the following theorem.
\begin{theorem}
The min-cost augmentation problem is NP-hard, even for strict lists of length
at most 3. Further, the lists can be derived from a master list.
\end{theorem}

\subsection{Inapproximability of min-cost augmentation}
\label{inapprox}
We extend the above reduction from $\I$ to show  that this problem is NP-hard to approximate
to within a factor of $\sqrt{n_1}/2$, where $n_1$ is the size of $\A$.
We construct a graph $H$ on at most $4m^4$ people
that satisfies the following property:
\begin{itemize}
\item[$(\ast)$] If $\I$ is a {\em yes} instance for 1-in-3 SAT, then $H$ can be augmented at a cost
of $m$ to admit a popular matching. If $\I$ is a {\em no} instance for 1-in-3 SAT,
then $H$ needs a cost strictly greater
than $m^3$ to admit a popular matching.
\end{itemize}
We describe the construction of the graph $H$ below.
Recall that $\I$ has $m$ clauses and corresponding to each clause $C_i$, we have a set $A_i$ of people.
The construction of $H$ is as follows. Let us call the group of 3 people $(a_4^i,a_5^i,a_6^i)$ in
Fig.~\ref{fig:clausegadget-mc-aug} a {\em triplet}. Instead of having just
one triplet in $A_i$, as was the case in the previous section, here we have many such triplets.
In particular, we have $m^3+1$ such triplets. The preference list
for one particular triplet $(a_{3t+1}^i, a_{3t+2}^i, a_{3t+3}^i)$ is shown in Fig.~\ref{fig:triplet}.
 \begin{figure}[ht]
 \centering
\begin{equation*}
  \setlength{\arraycolsep}{0.5ex}\setlength{\extrarowheight}{0.25ex}
  \begin{array}{@{\hspace{1ex}}c@{\hspace{1ex}}||@{\hspace{1ex}}c@{\hspace{1ex}}|@{\hspace{1ex}}c@{\hspace{1ex}}|@{\hspace{1ex}}c@{\hspace{1ex}}}
    a_{3t+1}^i & r_i^t \ & u_{j_1} & \\[.5ex] \hline
    a_{3t+2}^i & r_i^t \ & u_{j_2} & \\[.5ex] \hline
    a_{3t+3}^i & r_i^t \ & u_{j_3} & \\[.5ex]
\end{array}
\end{equation*}
\caption{Preference lists of people corresponding to the $t$-th triplet.}
\label{fig:triplet}
\end{figure}

We now have $3+3(m^3+1)$ people in $A_i$, namely
$a_1^i, a_2^i, a_3^i$ and 3 people per triplet, for each of the $m^3+1$ triplets. Thus our overall
instance $H$ has $m(3+3(m^3+1))$ (the people in $\cup_iA_i$), plus the $n$ people in $\{x_1,\ldots,x_n\}$.
Since each clause has 3 variables, $n \le 3m$. Thus we can bound
$n_1$, the number of people in $H$ as:
$n_1 \le 3m^4 + 9m \le 4m^4$ for $m \ge 3$.

Recall that for each $j$, the preference list of $x_j$ is of length 1, which consists of only $u_j$.
The costs of the items are as follows: the cost of each of the {\em internal} items, i.e., $p_i, q_i$, and $r_i^k$, for
$k = 1,\ldots,m^3+1$ is $m^3$, and the cost of each $u_j$ for $j = 1,\ldots,n$ is 1.
We now show that the instance constructed as above satisfies the property $(\ast)$.

\begin{lemma}
If $\I$ is a {\em yes} instance for 1-in-3 SAT, then $H$ can be augmented at a cost
of $m$ to admit a popular matching. If $\I$ is a {\em no} instance for 1-in-3 SAT,
then $H$ needs a cost strictly greater
than $m^3$ to admit a popular matching.
\end{lemma}
\begin{proof}
We first consider the case when $\I$ is an {\em yes} instance.
The proof is similar to that of Lemma~\ref{lemma5}.
For each $j$, where $1 \le j \le n$, do the following: if $X_j = true$, then set $\Capacity(u_j) = 1 + c_j$,
where $c_j$ is the number of clauses in which $X_j$ is present.
Else set $\Capacity(u_j) = 1$.
The total cost involved here is $\sum_{j: X_j = true} c_j$.
Since each clause has exactly one variable set to true, we have:
$\sum_{j: X_j = true} c_j = m$.
Thus, the cost of our instance $\tilde{H}$ is $m$.
It is easy to show that the graph $\tilde{H}'$ admits an $\A$-complete matching.
\begin{itemize}
\item Consider the people $x_1,\ldots,x_n$.
Each $x_j$ gets matched to her $f$-item $u_j$.
\item Consider the people in $A_i$. We know that
exactly one amongst $u_{j_1}, u_{j_2}, u_{j_3}$ has more than one copies
(since the number of copies  was based on a satisfying
assignment for 1-in-3 SAT). If $\Capacity(u_{j_k}) > 1$, then $a_k^i$ gets matched to $u_{j_k}$
and the 2 people in $\{a_1^i, a_2^i, a_3^i\}\setminus\{a_k^i\}$ get matched to 
$p_i$ and $q_i$.
For each of the $m^3+1$ triplets that we have here, we do as follows.
The person $a_{3t+k}^i$ gets matched to her top choice item $r_i^t$ whereas the 2 people in
$\{a_{3t+1}^i, a_{3t+2}^i, a_{3t+3}^i\}\setminus\{a_{3t+k}^i\}$
get matched to their last items.
\end{itemize}

This proves that $H$ can  be augmented at a cost of exactly $m$
to admit a popular matching.

\smallskip
We now prove the other direction, that is, if $\I$ is a {\em no} instance for 1-in-3 SAT,
then $H$ needs a cost of at least  $m^3+1$ to admit a popular matching.
Suppose $H$ can be augmented at a cost of at most $m^3$ to admit a popular matching. We will show that this
translates to a 1-in-3 satisfying assignment for $\I$.
Let $\tilde{H}$ denote the augmented graph. Let us set the truth values of variables in $\I$ as follows.
Set $X_j = true$ iff $\Capacity(u_j)$ in $\tilde{H}$ is greater than 1.

We have only
$m^3$ units available to make extra copies so that people in each set $A_i$ have items
in $\tilde{H}'$ to match  themselves to. Recall that the cost of each internal item is $m^3$. Hence
it is easy to see that we cannot afford an extra copy of any
internal item and thus at least one public item in $\{u_{j_1},u_{j_2},u_{j_3}\}$ should have more than one copy
to match all of $a_1^i,a_2^i,a_3^i$.
Otherwise there are only 2 items $p^i$ and $q^i$ for these 3 people
to be matched to; since the first copies of $u_{j_1},u_{j_2},u_{j_3}$ will be matched to $x_{j_1},x_{j_2},x_{j_3}$,
respectively. Thus, we have shown that at least one of $u_{j_1},u_{j_2},u_{j_3}$ has more than one copy.
Hence in our assignment of truth values, there is at least 1 variable in each clause that is set to true.

Suppose 2 or more of the items in $\{u_{j_1},u_{j_2},u_{j_3}\}$ have more than one copy in $\tilde{H}$.
We have two people in $\{a_1^i, a_2^i, a_3^i\}$ having their most preferred {\em even}
item in $\tilde{H}_1$ as
an item in $\{u_{j_1},u_{j_2},u_{j_3}\}$.
In addition, in each of the $m^3+1$ triplets, two people have their most preferred {\em even} item in
$\{u_{j_1},u_{j_2},u_{j_3}\}$.
Although, one of these 2 people from each triplet can be matched to her unique top choice item,
we still need to spend $m^3+1$ for all the people in $A_i$ to be matched to items in $\tilde{H}'$.
This contradicts the hypothesis  that $H$ can be augmented a cost of at most $m^3$ into $\tilde{H}$.
Hence for each $i$, there is exactly 1 item in $\{u_{j_1},u_{j_2},u_{j_3}\}$ that has more than one copy in
 $\tilde{H}$.
In other words, for each $i$, there is exactly 1 true variable in the $i$-th clause.
Thus, our assignment is a 1-in-3 satisfying assignment for $\I$.
\qed
\end{proof}

Now suppose that the min-cost augmentation problem admits a $\sqrt{n_1}/2$ approximation algorithm.
Call this algorithm $\Algo$. If $\I$ is a yes instance, then $\Algo$ has  to return an augmentation
of cost at most $1/2.\sqrt{4m^4}.m = m^3$. If $\I$ is a no instance, then there is no augmentation
of cost at most $m^3$, so $\Algo$ returns an answer of cost greater than $m^3$.
Thus using $\Algo$ it is possible to determine whether $\I$ has a 1-in-3 satisfying assignment or not,
a contradiction.
Hence we conclude the following theorem.

\begin{theorem}
It is NP-hard to approximate the min-cost
augmentation problem on $G = (\A\cup\B,E)$ within $\sqrt{|\A|}/2$.
\end{theorem}

\subsection{Perfect augmentation}
In this section we consider a variant of the min-cost augmentation problem
where we are not content with only a popular matching, but insist on every
applicant being matched to an item. Note that till now we allowed people
to be matched to their last items which was equivalent to leaving
them unmatched. We call a popular matching that matches all
the people as a {\em perfect popular matching} and
denote the problem as {\em min-cost perfect augmentation}.
We show that this problem becomes
NP-hard even when preference lists are strict and have length at most $2$ in contrast to the min-cost
augmentation problem which has a simple polynomial time algorithm for
strict lists of length at most $2$.

The overall reduction is similar to the min-cost augmentation
problem where we reduce
from an instance $\I$ of monotone 1-in-3 SAT problem.
As earlier, corresponding to every variable $X_j$ in $\I$
we have an applicant $x_j$ in our instance. The preference
list of applicant $x_j$ contains only one item $u_j$.
Let $C_i = (X_{j_1} \vee X_{j_2} \vee X_{j_3})$ be a clause in $\I$.
Corresponding to clause $C_i$ we have a set of 6 people
$A_i = \{ a_1^i, \ldots, a_6^i\}$ and a set of internal
items $D_i = \{p_i, q_i\}$. The internal items
appear on preference lists of people only in
$A_i$. In addition public items $u_{j_1}, u_{j_2}, u_{j_3}$
also appear on the preference lists of people in
$A_i$. The preference lists of people
are as shown in Fig.~\ref{fig:clausegadget-min-cost-perf-aug}.

 \begin{figure}[ht]
\begin{minipage}[b]{0.44\linewidth}
 \centering
\begin{equation*}
  \setlength{\arraycolsep}{0.5ex}\setlength{\extrarowheight}{0.25ex}
  \begin{array}{@{\hspace{1ex}}c@{\hspace{1ex}}||@{\hspace{1ex}}c@{\hspace{1ex}}|@{\hspace{1ex}}c@{\hspace{1ex}}|@{\hspace{1ex}}c@{\hspace{1ex}}|@{\hspace{1ex}}c@{\hspace{1ex}}|@{\hspace{1ex}}c@{\hspace{1ex}}}
    a_{1}^i & p_i \ & u_{j_1} \ \\[.5ex] \hline
    a_{2}^i & p_i \ & u_{j_2} \ \\[.5ex] \hline
    a_{3}^i & p_i \ & u_{j_3} \ \\[.5ex] 
\end{array}
\end{equation*}
\end{minipage}
\begin{minipage}[b]{0.44\linewidth}
\centering
\begin{equation*}
  \setlength{\arraycolsep}{0.5ex}\setlength{\extrarowheight}{0.25ex}
  \begin{array}{@{\hspace{1ex}}c@{\hspace{1ex}}||@{\hspace{1ex}}c@{\hspace{1ex}}|@{\hspace{1ex}}c@{\hspace{1ex}}|@{\hspace{1ex}}c@{\hspace{1ex}}}
    a_{4}^i & u_{j_1} \ & q_i & \\[.5ex] \hline
    a_{5}^i & u_{j_2} \ & q_i & \\[.5ex] \hline
    a_{6}^i & u_{j_3} \ & q_i & \\[.5ex] 
\end{array}
\end{equation*}
\end{minipage}
\caption{Preference lists of the 6 people in $A_i$}
\label{fig:clausegadget-min-cost-perf-aug}
\end{figure}

We define the
sets $\A$ and $\B$ below.
\begin{eqnarray*}
\A &=& \cup_{i=1}^m A_i \cup \{x_1,\ldots,x_n\}.\\
\B &=& \cup_{i=1}^m D_i \cup \{u_1,\ldots,u_n\}.
\end{eqnarray*}
The graph $G=(\A \cup \B, E)$ with the preference lists as defined above is
our instance $G$. It remains to describe the costs of items in $G$.
The cost of every public item $u_j$ is 1 whereas the costs
of every internal item is $m$.
With a single copy of every item, all public items are critical in
the graph $G_1$. Hence people $a_1^i, a_2^i, a_3^i$ belonging to $A_i$ treat their
unique last resort items as their $s$-items. Further people $a_4^i, a_5^i, a_6^i$
treat the internal item $q_i$ as their $s$-item.
It is easy to verify that the graph $G$ described above does not admit a
popular matching.

Let $\tilde{G}$ be an augmented instance that admits a perfect popular
matching. In any perfect popular matching in $\tilde{G}$ every applicant
$x_j$ has to be matched to her $f$-item i.e. $u_j$.
Now let us consider the people in $A_i$. Since the internal  items
are of cost $m$ each and public items have cost 1, it is cheaper to
make extra copies of public items.
Further, recall that any popular matching has to be a maximum cardinality matching on rank-1 edges.
Therefore in order to match one of $\{a_1^i, a_2^i, a_3^i\}$ (say $a_1^i$) to
its $s$-item $u_{j_1}$, we need to make 2 extra copies of $u_{j_1}$
thus making
$u_{j_1}$ non-critical in $\tilde{G}_1$. This ensures that applicant $a_4^i$ gets
matched to a copy of $u_{j_1}$, her $f$-item.
It is easy to see that in order to get a perfect popular matching,
we need to make at least two public items
non-critical in the graph $\tilde{G}$.
Thus we spend an amount of at least $4$ units per clause
in any augmented instance $\tilde{G}$ that admits a perfect popular matching.

We now show the following lemma which proves the correctness of our reduction.
\begin{lemma}
The min-cost instance $\tilde{G}$ corresponding to $G$ that admits a perfect
popular matching has cost $4m$ iff there exists a 1-in-3 satisfying assignment
for $\I$.
\end{lemma}
\begin{proof}
Assume that there exists a 1-in-3
satisfying assignment for $\I$ and let $Truthval$ denote the 1-in-3 satisfying assignment for $\I$.
For every variable $X_j$, let $c_j$ denote the number of clauses in which $X_j$
appears.
We set the copies of the items in $\tilde{G}$ as follows:
If  $Truthval(X_j) = false$, then $\Capacity(u_j) = 2c_j + 1$,
otherwise $\Capacity(u_j) = 1$.
Thus we pay $2c_j$ units for the extra copies of item $u_j$ whenever the corresponding
variable $X_j$ is set to false.

Since $\I$ has a 1-in-3 satisfying assignment, each clause has exactly two variables set to false.
We therefore have:
\begin{eqnarray*}
\sum_{j: Truthval(X_j) = false} 2c_j = 4m.
\end{eqnarray*}
Thus the cost of our instance $\tilde{G}$ is $4m$.
We now show that in the graph $\tilde{G}$, every applicant has an item to match amongst her $f$ or $s$ items.
\begin{itemize}
\item Consider the people $x_1, x_2,\ldots, x_n$.
Each $x_j$ gets matched to her $f$-item $u_j$.
\item Consider the people in $A_i$ corresponding to
the clause $C_i = (X_{j_1} \vee X_{j_2} \vee X_{j_3})$. We know that
exactly two among $u_{j_1}, u_{j_2}, u_{j_3}$ are non-critical. Assume
that $u_{j_1}$ and $u_{j_2}$ are non-critical, then people  $a_1^i$ and $a_2^i$
get matched to their $s$-items $u_{j_1}$ and $u_{j_2}$ respectively. Further
$a_4^i$ and $a_5^i$ get matched to their $f$-items $u_{j_1}$ and $u_{j_2}$ 
respectively.
Finally, $a_3^i$ gets matched to her $f$-item $p_i$ whereas $a_6^i$
gets matched to her $s$-item $q_i$.
\end{itemize}

\medskip
To prove the other direction, assume that $\tilde{G}$ admits a
popular matching and has cost $4m$.
We now translate this into the truth values for variables in $\I$.
If $\Capacity(u_j) > 1$ in $\tilde{G}$, then set variable $X_j = false$ else
set $X_j = true$.
We need to show that this assignment is a 1-in-3 satisfying assignment.

Recall that for every clause we need to spend at least 4 units in order to
match all the people in $A_i$. Since the cost of $\tilde{G}$ is
$4m$, this implies that exactly $4$ units have been spent per clause for making
extra copies.

Now consider the people in $A_i$. The internal items occurring on
their preference lists of these people have cost $m$ each, hence
none of the internal items have extra copies in the augmented instance.
This implies that at least 2 public items amongst $\{u_{j_1}, u_{j_2}, u_{j_3}\}$
have more than one copy in the augmented instance. Thus our truth assignment
sets at least 2 variables to $false$. It remains to prove that for every
clause exactly 2 variables are set to $false$. Assume not.

Let $C_i$ be a clause in which our truth assignment set all the three variables
$X_{j_1}, X_{j_2}, X_{j_3}$ to $false$. This implies that in the instance $\tilde{G}$
all the three public items $u_{j_1}, u_{j_2}, u_{j_3}$ have more than one copy.
The first copy of these public items is utilized in matching people $x_{j_1}, x_{j_2},
x_{j_3}$ respectively to their $f$-items. Further since any popular matching
is a maximum cardinality matching on rank-1 edges, the second copies
of these items will have to be assigned to people $a_4^i, a_5^i, a_6^i$.
Thus after spending 3 units we are still left with 2 people amongst
$\{a_1^i, a_2^i, a_3^i\}$ without any item. Hence we need to spend 2 more units
to match 2 of these people, which implies that we spent 3+2=5 units
for people in $A_i$. This contradicts the fact that exactly $4$ units
was spent in order to match people in $A_i$, for all $i$.
Thus exactly one variable in every clause is set to $true$.
\qed
\end{proof}
The following theorem is immediate from the above lemma.
\begin{theorem}
The min-cost perfect augmentation problem is NP-hard for strict lists of length
at most $2$.
\end{theorem}

\section{Min-cost Popular matchings}
\label{min-cost-pop-mat}
In this section we present an $O(mn_1)$ time algorithm for the min-cost popular matchings problem, where $m = |E|$ and $n_1 = |\A|$.
Our input is an instance $G = (\A \cup \B, E)$ where each item $b \in \B$ has associated with it the number $\Capacity(b)$  
(denoting the maximum number of people that can be matched to $b$) and a price $\cost(b) \ge 0$. 
Whenever a person gets matched to $b$, an amount of $\cost(b)$ has to be paid. Thus if $k \le \Capacity(b)$ copies of $b$ gets
used in a matching $M$, then a cost of $k\cdot \cost(b)$ has to be paid by $M$. 
As done in the earlier sections, we will add a last item $\ell_a$ at the end of $a$'s preference
list for each person $a \in \A$. 
The cost of $\ell_a$ is 0, since using the edge $(a,\ell_a)$ amounts to leaving
$a$ unmatched.

Our problem here is to decide whether $G$ admits a popular matching or not and 
if so, to compute the one with minimum cost.
As mentioned in Section~\ref{intro},
Manlove and Sng considered the popular matchings problem (referred to as the CHAT problem)
where items (these were called houses) have capacities and 
they showed an $O((\sqrt{C}+n_1)m)$ algorithm for this problem, where 
$C$ is the sum of all the capacities. 

In order to solve the min-cost popular matchings problem, for each $b \in \B$, we could make 
$\Capacity(b)$ copies of each vertex $b$, call them $b_1,\ldots,b_{\Capacity(b)}$, 
where each $b_i$ has the same neighborhood as the original vertex $b$.
However, such a graph has too many vertices and edges,  
hence we will stick to the original graph $G = (\A\cup\B,E)$ and simulate the larger graph in $G$ itself.
Note that a {\em matching} in $G$ can contain up to $\Capacity(b)$ pairs $(a_i,b)$. 
It is easy to see that the structural characterization for popular matchings from \cite{AIKM05} holds
for our problem as well. That is, any 
popular matching in our graph $G$ has to be a maximum cardinality matching on rank-1 edges and 
every person $a$ has to  be matched to an item in $f(a)\cup s(a)$.
This is because by making $\Capacity(b)$ many duplicates of every item $b$ in $G$ our problem becomes equivalent to 
the original popular matchings problem.

\subsection{Our algorithm}
Our algorithm to compute a min-cost popular matching can be broadly partitioned
into two stages. In the first stage we build the graph $G'$, i.e. the graph
where every person adds edges to their $f$ and $s$-items. Identifying
$s$-items for people involves partitioning the vertices of $G$
into {\em odd, even} and {\em unreachable} with respect to
a maximum cardinality matching on rank-1 edges. We show how to efficiently
do this by building Hungarian trees rooted at unmatched vertices.
The second stage then computes a min-cost popular matching in the graph
$G'$ if one exists.

\subsubsection{The first stage.}
We first construct the graph $G_1$ which is the graph $G$ restricted to rank-1 edges.
In order to find a maximum cardinality matching in the graph $G_1$, we use Ford-Fulkerson max-flow
algorithm. The following transformation from $G_1$ into a flow network
is based on the standard transformation from the bipartite matching problem to the
maximum flow problem:
\begin{itemize}
\item add a vertex $s$ and an edge directed from $s$ to each person $a\in\A$ with an edge capacity of 1 on this edge.
\item add a vertex $t$ and an edge directed from each item $b\in\B$ to $t$ with an edge capacity of $\Capacity(b)$ on this edge.
\item direct every edge $(a,b)$ of $G$ from $a$ to $b$ and set an edge capacity of 1 for each such edge.
\end{itemize}
Let $F(G_1)$ denote the above graph. It is easy to see that a valid flow from $s$ to $t$
in the graph $F(G_1)$ can be translated to
a {\em matching} in $G_1$ in which every person is matched to at most 1 item
and every item $b$ is  matched up to $\Capacity(b)$ people. A maximum flow in $F(G_1)$ becomes a maximum
cardinality matching in $G_1$.
We compute a maximum cardinality matching $M_0$ of  $G_1$ by computing a max-flow from $s$ to $t$ in $F(G_1)$.
Using the matching $M_0$, our goal is to obtain a partition of $\A\cup\B$ into $\odd$ ({\em odd}), $\E$ ({\em even}) and
$\U$ ({\em unreachable}).
This can be done in $O($number of edges$)$ provided we create
$\Capacity(b)$ many duplicates of each item $p$ and duplicate the neighborhood of $b$ for each copy of $b$.
However this is too expensive. The main point to note is that all the
$\Capacity(b)$ many copies of $b$, for each item $b$, have the same {\em odd/even/unreachable} status.
We show below that we can remain in the
graph $G_1$ and determine the {\em odd/even/unreachable} status of all the vertices in linear time.

\begin{enumerate}
\item We begin with $\odd = \E = \U = \emptyset$.
\item We then add to the set $\E$ all people
that are unmatched in $M_0$ and all items that are not fully matched by $M_0$ (i.e. a item $b$ that is
matched to fewer than $\Capacity(b)$ many people). This is because if we would have made $\Capacity(b)$ many
duplicates of $b$, some of the copies would have remained unmatched by $M_0$ and the other copies which are matched
would be connected by {\em even} length alternating paths from these unmatched vertices.

\item Our goal now is to build a Hungarian tree $T_u$ for each vertex $u$
that is unmatched or not fully in $M_0$. In
order to do so we first set all vertices as unmarked. We build the trees rooted at 
unmatched people
and not fully matched items as described below:

\begin{enumerate}
\item For $u \in \A$ that is unmatched, the children of $u$ in $T_u$ are all the neighboring items of $u$ that are unmarked so far.
For each of these items $b$ the children of $b$ in $T_u$ are all the unmarked people matched to $b$.
The children of these people are their neighboring unmarked items and so on. As soon
as a vertex gets visited in $T_u$ we mark it.

\item For $u \in \B$ the children of $u$ are all the neighboring unmarked people of $u$.
Note that some of these people could be matched to $u$ -- however, we will include 
all these people
since we are simulating the Hungarian tree rooted at an {\em unmatched} copy of $u$. We mark
each person in this child list.

Each person $a$ in the above child list had a unique child, the item to which $a$ is matched.
If this item is marked, then $a$ is a leaf in this tree, else we add $M_0(a)$ to the tree and mark
it. We now continue to explore the unmarked neighborhood of $M_0(a)$ for all non-leaf people $a$.

\item Once $T_u$ is built, all vertices that belong to {\em even} levels of $T_u$ (the root is at level 0)
are added to $\E$ and all
vertices that belong to {\em odd} levels are added to $\odd$.

\end{enumerate}
\item Once we finish building all the trees $T_u$, where $u$ is unmatched person/not a fully matched item,
the set $\U$ gets set to the vertices of $\A\cup\B \setminus \odd\cup\E$ as there is no alternating path from
an unmatched vertex to such vertices.

\end{enumerate}

We note that while building a tree $T_u$, we explore the neighborhood of a vertex only if this
vertex is {\em unmarked} and then this vertex immediately gets marked. This ensures that
a vertex occurs just once across all $T_u$'s.
Having obtained the partition, it is now possible to define $s(a)$ for every
person $a$ as the most preferred {\em even} item of $a$.
Let the graph
$G'$ be the graph $G_1$ along with the edges $(a,b)$ where $a\in\E$ and $p \in s(a)$.

Since a popular matching is a maximum cardinality matching on
rank-1 edges, all items that are {\em critical} in $G_1$, that is, all items in $\odd\cup\U$ have
to be fully matched in every popular matching $M^*$ of $G$. The only choice we have is in choosing which
items of $\E$ should participate with how many copies
in the min-cost popular matching. We make this choice in the
second stage of our algorithm.

\subsubsection{The second stage.}
Our goal in the second part of the algorithm is to augment the matching $M_0$ to find a min-cost popular matching.
However, we start with the matching $M_1$, where $M_1 = M_0 \setminus \{$all edges $(a, b)$ where $a \in \odd$\}.
Thus $M_1$ consists only of edges $(a,b)$ where $b \in \odd \cup \U$.
We take $M_1$ to be our starting matching rather than $M_0$ because it may be possible
to match people $\odd \cap \A$ to cheaper rank-1 neighbors. Recall that while computing
the max-flow $M_0$, the costs of items played no role.

Now let $\rho$ be an augmenting path with respect to $M_1$, i.e., one end of $\rho$ is an unmatched person
and the other end of $\rho$ is item $b$ that is not fully matched.
The cost of augmenting the current matching along $\rho$ is the cost of $b$.
By augmenting the current matching along $\rho$,
every item other than $b$ that is currently matched stays matched to the same number of people
 and the item $b$
gets matched to one more person. Thus the cost of the new matching
is the cost of the old matching $+$ $\cost(b)$. In order to match an unmatched person $a$,
our algorithm always chooses the cheapest augmenting path starting from the person $a$.

To find the cheapest augmenting path we build a Hungarian tree $T_a$ rooted at every
person that is unmatched in $M_1$.  Initially all vertices are unmarked and while building $T_a$
every visited vertex gets marked so that each vertex occurs at most once in $T_a$. We do not
terminate the construction of $T_a$ as soon as we find an augmenting path, but we build $T_a$
completely in order to find a min-cost item $b$ such that there is an augmenting path between and $a$
and $b$; we augment $M_1$ along this path to obtain $M_2$. On the other hand if $T_a$ has no augmenting path then we quit and declare ``$G$ does
not admit a popular matching".

\begin{algorithm}
\begin{algorithmic}[1]
\STATE Construct the graph $G_1 = (\A \cup \B, E_1)$ where $E_1 = \{(a,b): a\in\A, b \in f(a)\}$.
\STATE Construct the flow graph $F(G_1)$ by adding two vertices $s$ and $t$
and adding directed edges with appropriate capacities.
\STATE Compute a maximum flow in $F(G_1)$ and translate the flow to a
matching $M_0$ in $G_1$.
\STATE Obtain a partition of the vertices of $G$ as {\em odd} ($\odd$) , 
{\em even} ($\E$)
and {\em unreachable} ($\U$) using $M_0$.
\STATE Construct the graph $G' = (\A \cup \B, E')$ where every person
adds edges to her $f$-items and every {\em even} person adds edges to
her $s$-items.
\STATE Delete from $G'$ all $\odd \odd$ and $\odd \U$ edges.

\STATE Delete from $M_0$ all edges that are incident on {\em odd} people
in $G'$ and call the resulting matching $M_1$.

\STATE $i=1$.
\WHILE {there exists an unmatched person $a$ in $M_i$}
\STATE Build a Hungarian tree $T_a$ rooted at $a$.
\IF {there exists no augmenting path starting at $a$}
\STATE Quit and declare ``$G$ does not admit any popular matching".
\ELSE
\STATE Augment $M_i$ along the cheapest augmenting path in $T_a$ and call
the new matching $M_{i+1}$.
\ENDIF
\STATE $i = i+1$.
\ENDWHILE
\STATE Return $M_i$.
\end{algorithmic}
\caption{Algorithm for min-cost popular matching.}
\label{algo:min-cost-popular}
\end{algorithm}

We present our entire algorithm in Algorithm~\ref{algo:min-cost-popular}.
To see the correctness of the algorithm we first note that
if there is no augmenting path in $T_a$, where $a$ is an unmatched person
in $M_i$, then there is no popular matching in $G$.
This is because every popular
matching is a maximum cardinality matching on  rank-1 edges and has to match every $a \in \A$
to a item in $f(a) \cup s(a)$.
It remains to prove that if $G$ admits a popular matching, then
the matching $M (=M_i)$ returned at Step 18 of Algorithm~\ref{algo:min-cost-popular} is a min-cost popular matching.
We prove that using Lemma~\ref{lemma6}.
\begin{lemma}
\label{lemma6}
If $G$ admits a popular matching, then
the matching $M$ returned by our algorithm is a min-cost popular matching in $G$.
\end{lemma}
\begin{proof}
Suppose $M$ is not a min-cost popular matching in $G$ and let $\OPT$ be such a matching.
For the purpose of this proof we operate on the {\em cloned} graph where each item $b$
has $\Capacity(b)$ many copies and $M$ and $\OPT$ both refer to matchings where each
item is matched to at most one person.
Consider $\OPT \oplus M$ - this is a collection of cycles and even length alternating paths
(since both $\OPT$ and $M$ are $\A$-complete). The cycles do not contribute to any change in 
costs since both $\OPT$ and $M$ match the same items in any cycle.

Let $\rho$ be a path in $\OPT \oplus M$. Let $\beta_0$ and $\beta_M$ be the endpoints of this path, where
$\OPT$ leaves $\beta_M$ unmatched while $M$ leaves $\beta_0$ unmatched. It suffices to show that
$\cost(\beta_M) \le \cost(\beta_0)$. Since $\OPT$ is a popular matching, it has to
match all the items in $\odd\cup\U$ (the odd/unreachable items in $G_1$). Since it leaves
$\beta_M$ unmatched, it follows that $\beta_M \in \E$ and thus there are items of $\E$ in $\rho$.

It is the second stage of
our algorithm that matches items in $\E$. 
Let $\alpha_1$ be the last person in the path $\rho$ to get matched by our algorithm
and let $M(\alpha_1) = \beta_1$. Since $\beta_0$ is unmatched in $M$ it implies that
during the execution of our algorithm we found at least two augmenting paths from
$\alpha_1$ -- one ending in $\beta_1$ and the other ending in $\beta_0$. Further,
we found the augmenting path ending in $\beta_1$ cheaper, that is, $\cost(\beta_1) \le \cost(\beta_0)$.

We now repeat the same argument for the $\beta_1$-$\beta_M$ sub-path of $\rho$. Let $\alpha_2$
be the last person in the  $\beta_1$-$\beta_M$ sub-path that got matched by our algorithm and
let $M(\alpha_2) = \beta_2$. Note that $\beta_1$ was also unmatched at this time and hence
our algorithm found at least two augmenting paths from $\alpha_2$ -- one ending in $\beta_1$
and another ending in $\beta_2$. Since $M(\alpha_2) = \beta_2$ it implies that
$\cost(\beta_2) \le \cost(\beta_1)$.

Repeating the same argument for the $\beta_2$-$\beta_M$ sub-path  we get vertices $\beta_3,\ldots,
\beta_t=\beta_M$ where $\cost(\beta_2) \le \cost(\beta_3) \le \cdots \le \cost(\beta_t)$.
Combining all the inequalities yields $\cost(\beta_M) \le \cost(\beta_0)$. \qed
\end{proof}

\paragraph{Time complexity of this algorithm.}
The difference between our algorithm and 
that of Manlove and Sng for the CHAT problem in the first stage is that
they use Gabow's algorithm to find a matching on rank-1 edges 
whereas we use Ford-Fulkerson max-flow algorithm. 
Gabow's algorithm runs in time $O(\sqrt{C}m)$ where $C = \sum_{i=1}^{|\B|} \Capacity(b_i)$ whereas
since the value of max-flow in the graph
$F(G_1)$ is upper bounded by $|\A| = n_1$, Ford-Fulkerson algorithm takes
$O(mn_1)$ time. Also, the total time taken by our algorithm to partition vertices into
$\odd, \E$, and $\U$ is $O(m+n)$ where $n$ denotes the total number of vertices in $G$.
It is easy to see that the time spent by our algorithm in the second stage is also $O(mn_1)$ 
since it takes $O(m)$ time to build the tree $T_a$ and there are at most $n_1$ such trees
that we build.
We can now conclude the following theorem.
\begin{theorem}
There exists an $O(mn_1)$ time algorithm to decide whether a given instance
$G$ of the min-cost popular matchings problem admits a popular matching and if so, to
compute one with min-cost.
\end{theorem}

Note that by assigning a huge cost $\hat{C} > \sum_b \Capacity(b).\cost(b)$ to each of the last items $\ell_a, a \in \A$ 
that we introduced, our algorithm also works for the min-cost maximum-cardinality popular matching problem where we
seek among all popular matchings of maximum cardinality, the one with minimum cost.

\subsection*{Conclusions}
In this paper we considered several extensions of the popular matching problem.
We showed that the min-cost popular instance problem, which involves
building a min-cost graph that admits a popular matching that matches
all applicants, is NP-hard,
even when preference lists are strict and of length at most 2. In contrast, the
min-cost {\em augmentation} problem admits a simple polynomial time algorithm
when preference lists are strict and of length at most 2. However, the
min-cost augmentation problem is
NP-hard in general; it is NP-hard 
even when preference lists are strict and of length at most 3. In fact,
it is NP-hard to approximate
the min-cost augmentation problem to within a factor of $\sqrt{n_1}/2$,
where $n_1$ is the number of people.
We also showed that the min-cost popular matching problem (the number
of copies of each item is fixed here) can be solved
in $O(mn_1)$ time, where $m$ is the number of edges in the input graph.

\bibliographystyle{abbrv}
\bibliography{references}

\end{document}